\documentclass[doublecol]{epl2}

\usepackage{color}

\newcommand{\correction}[1]{{#1}}

\title{Detecting unstable periodic spatio-temporal states \\ of spatial extended chaotic systems}
\shorttitle{Detecting UPSTSs} 

\author{Alexander E.~Hramov\inst{1} \and Alexey A.~Koronovskii\inst{1}}
\shortauthor{A.E.~Hramov\etal}

\institute{
  \inst{1} Faculty of Nonlinear Processes, Saratov State
University - Astrakhanskaya, 83, Saratov, 410012, Russia
}


\usepackage{ams}

\abstract{The method of detection of the unstable periodic
spatio-temporal states of spatial extended chaotic systems has been
proposed. The application of this method is illustrated by the
consideration of two different systems: (i) the fluid model of
Pierce diode being one of the fundamental system of the physics of
plasmas and microwave electronics and (ii) the complex
one-dimensional Ginzburg-Landau equation demonstrating different
regimes of spatio-temporal chaos.}

\begin{document}

\maketitle


It is well known that the unstable periodic orbits (UPOs) embedded
into chaotic attractors play an important role in the dynamics of
the systems with a small number of the degree of
freedom~\cite{Cvitanovic:1988_cycles,%
Kostelich:1989_experiment,Barreto:1997_bifurcations}. The chaotic
regime of the system may be characterized by means of the set of
UPOs~\cite{Carroll:1999_UnstableOrbits}. A universal and powerful
tool for exploration of chaotic
dynamics~\cite{Cvitanovic:1991_orbits}, unstable periodic orbits
proved  to be especially efficient in the context of chaotic
synchronization \cite{Rulkov:1996_SynchroCircuits,
Pikovsky:1997_EyeletIntermitt, Pikovsky:1997_PhaseSynchro_UPOs,
Aeh:2005_SpectralComponents}. The different types of chaotic
synchronization (such as phase
synchronization~\cite{Pikovsky:1997_PhaseSynchro_UPOs},
lag~\cite{Rosenblum:1997_LagSynchro} and complete
synchronization~\cite{Pikovsky:1991_CSSymmetryBreaking}) may be
explained in terms of unstable periodic
orbits~\cite{Pazo:2002_UPOsSynchro}. Eventually, UPOs play the key
role for the chaos controlling
problem~\cite{Bielawski:1993_ControllingChaos} since the unstable
periodic orbits may be stabilized by means of the week influence
on the system dynamics, e.g., with the help of small variation of
the control parameter~\cite{Ott:1990_ControllingChaos} or with the
feedback of different types~\cite{Pyragas:1992_ControllingChaos,
Chen:1994_ControllingChaos, Pierre:1996_ControlChaosPlasma}.

In the spatial extended systems the unstable periodic
spatio-temporal states (UPSTSs) exist~\cite{Franceschini:1999} which
are similar to the unstable periodic orbits in the chaotic systems
with a small number of the degree of freedom. In particular, the
chaotic dynamics of spatial extended systems may be controlled by
stabilizing such unstable periodic spatio-temporal
states~\cite{Boccaletti:1999_ControllingChaos}. Therefore, one of
the important problem connected with the study of the spatial
extended chaotic system is finding these unstable periodic states.
It is appropriate to suggest that the methods aimed at the search of
UPOs of discrete maps (and the flow dynamical systems with small
dimension of phase space, too) may be adapted to the spatial
extended systems. The method proposed by D.P.\,Lathrop and
E.J.\,Kostelich \cite{Kostelich:1989_experiment}, as an example, had
been used to pick out UPSTSs for the fluid model of Pierce
diode~\cite{Rempen:2004_PierceDiode}. This method is based on the
obtaining the histograms describing the frequency of system
returning to the vicinity of UPOs (in the systems with a small
number of the degree of freedom) or UPSTSs (in the spatial extended
systems), respectively. Nevertheless, this method applied to spatial
extended systems is rather imprecise and time-consuming. Let us also
note the work of S.M.\,Zoldi and H.S.\,Greenside
\cite{Zoldi:1998_UPOs}, where the numerical analysis of UPOs for a
high-fractal-dimension chaotic solution of the partial differential
equation is carried out with the help of the innovative
damped-Newton method.

In this Letter we describe the modification of the method of
P.\,Schmelcher and F.\,Diakonos (SD--method)
\cite{Schmelcher:1997_UnstableOrbit,Pingel:2001_SD-methodPRE}
allowing precise detection of UPSTSs in the spatial extended chaotic
systems. As the sample analyzed spatially extended chaotic systems
we consider here the fluid model of Pierce diode and the complex
one-dimensional Ginzburg-Landau equation.

As the primary system under study we have used the fluid model of
Pierce diode \cite{Godfrey:1987, Kuhn:1990, Lindsay:1995,
Matsumoto:1996, Hramov:2004_IJE, Filatov:2006_PierceDiode_PLA} being
one of the simplest beam-plasma systems demonstrating complex
chaotic dynamics. It consists of two plane parallel infinite grids
pierced by the monoenergetic (at the entrance) electron beam. The
grids are grounded and the distance between them is $L$. The
entrance space charge density $\rho_0$ and velocity $v_0$ are
maintained constant. The space between the grids is evenly filled by
the neutralizing ions with density $|\rho_i/\rho_0|=1$. The dynamics
of this system is defined by the only parameter, the so-called
Pierce parameter $\alpha={\omega_pL}/{v_0}$, where $\omega_p$ is the
plasma frequency of the electron beam. With $\alpha>\pi$ in the
system, the so-called Pierce instability \cite{Pierce:1944,
Matsumoto:1996} develops, which leads to the appearance of the
virtual cathode. At the same time, with $\alpha\sim3\pi$, the
instability is limited by non-linearity and the regime of complete
passing of the electron beam through the diode space can be
observed. In this case the system can be described by the partial
differential equations:
\begin{equation}\label{eq:PDE}
\displaystyle\frac{\partial v}{\partial t}+v\frac{\partial
v}{\partial x}=\frac{\partial\varphi}{\partial x}, \ \ \ \ \
\displaystyle\frac{\partial\rho}{\partial t}+ v\frac{\partial
\rho}{\partial x}+\rho\frac{\partial v}{\partial x}=0, \\
\end{equation}
\begin{equation}\label{eq:PDE_Puasson}
\displaystyle\frac{\partial^2\varphi}{\partial
x^2}=\alpha^2(\rho-1),
\end{equation}
with the boundary conditions:
\begin{equation}\label{eq:BounCond}
v(0,t)=1,\quad \rho(0,t)=1,\quad\varphi(0,t)=\varphi(1,t)=0.
\end{equation}



In equations (\ref{eq:PDE}) the non-dimensional variables (space
charge potential $\varphi$, density $\rho$, velocity $v$, space
coordinate $x$ and time $t$) are used. They are related to the
corresponding dimensional variables as follows:
\begin{equation}\label{q6}
\begin{array}{c}
  \varphi '=  ({v^2_0/\eta})\varphi,\quad E'=({v^2_0/L\eta})E,
  \\
   \rho' =\rho_0\rho,\quad v'=v_0 v,\quad x'=L x,\quad t'=({L/v_0})t, \\
\end{array}
\end{equation}
where the dotted symbols correspond to the dimensional values,
$\eta$ is the specific electron charge, $v_0$ and $\rho_0$ are the
non-perturbed velocity and density of the electron beam, $L$ is the
length of the diode space. Equations (\ref{eq:PDE}) are integrated
numerically with the help of the one-step explicit two-level scheme
with upstream differences and Poisson equation
(\ref{eq:PDE_Puasson}) is solved by the method of the error vector
propagation. The time and space integration steps have been taken as
$\Delta t=0.003$ and $\Delta x =0.005$, respectively.

One of the core problems related to the spatial extended system
consideration is the infinite dimension of the ``phase space''
$W^{\infty}$. As a consequence, the state  ${\bf U}(x,t)$ of the
system of investigation should be considered instead of vector
$\mathbf{x}(t)$ in $\mathbb{R}^n$ as in the case of the flow
systems\footnote{For the system under consideration~(\ref{eq:PDE})
this state ${\bf U}(x,t)$ is the vector of the functions
characterizing the system dynamics, i.e., ${\bf U}(x,t)=(v(x,t),
\rho(x,t), \varphi(x,t))^T$.}. After the transient finished (i.e.,
$t>t_{tr}$) the set of the states ${\bf U}(x,t)$, $\forall t>t_{tr}$
may be considered as attracting subspace $W^s$ of the
infinite--dimensional ``phase space'' $W^{\infty}$ of the spatial
extended system under study. If the dimension of this subspace is
finite, the finite-dimensional space $\mathbb{R}^m$ of variables
may be used to describe the dynamics of the spatial extended system.

In is well-known that SD-method was developed to the UPOs detection
in the systems with discrete time, although it may be also applied
to the flow systems~\cite{Pingel:2001_SD-methodPRE} by means of
reducing them to maps with the help of Poincar\'e secant.
\correction{In order to apply the SD method to an extended system,
we assume that its infinite-dimensional phase space possesses the
low-dimensional attracting invariant subspace $W^s$, and the desired
solution lies in this subspace. Further, we construct the auxiliary
system $\mathbf{y}(t)$ in which the vector field $\mathbf{y}$ is in
one-to-one correspondence with $W^s$.}


\correction{The stationary states ${\bf U}^0(x,t)={\bf U}^0(x)$ of
the spatial extended system correspond to the fixed points in the
phase space of the auxiliary system, while the periodic
spatio-temporal states of~(\ref{eq:PDE}), (\ref{eq:PDE_Puasson}) are
in one-to-one correspondence with the periodic orbits of the
finite-dimensional system $\mathbf{y}(t)$. Therefore, UPSTSs of
spatial extended system may be found by means of the detection of
UPOs of the auxiliary finite-dimensional system}.

There are many well-known methods for applying low-dimensional
variable space to describe the behavior of the spatial extended
system, among which a typical one is the mode expansion method. In
particular, in Ref.~\cite{Hramov:2004_IJE} the low-dimensional model
has been constructed for Pierce
diode~(\ref{eq:PDE})--(\ref{eq:PDE_Puasson}) by means of the
extraction of several principal modes with the help of Galerkin
method. In the present work we propose the use of the variables
taken from several points $x_i$ of the extended system space to
construct the finite dimensional system
\begin{equation}
{\mathbf{y}(t)=(\rho(x_1,t),\dots,\rho(x_m,t))^T},
\label{eq:FlowFromPDE}
\end{equation}
where $m$ is the dimension of the auxiliary system,
${x_i=iL/(m+1)}$, $i=\overline{1,m}$. In comparison with the other
known methods, such approach allows us to undergo easily from the
spatial extended system state $\mathbf{U}(x,t)$ to the
low--dimensional vector $\mathbf{y}(t)$ without any additional
calculations or measurements.


For the system under study~(\ref{eq:PDE})--(\ref{eq:PDE_Puasson}) we
have estimated the dimension of the auxiliary vector $\mathbf{y}(t)$
as $m=3$. This assumption is based on the previous results of the
consideration of the finite-dimensional model of the Pierce diode
dynamics obtained with the help of Galerkin
method~\cite{Hramov:2004_IJE}. 

To confirm meeting of the requirements of the one-to-one
correspondence between state ${\bf U}(x,t)$ of the spatial extended
system and vector $\mathbf{y}(t)$ of the constructed auxiliary
system with the small number of degree of freedom we have used the
neighbour method \cite{Pecora:1995_statistics}. We have examined
that the distance
$d(\mathbf{y}_1,\mathbf{y}_2)=||\mathbf{y}_1-\mathbf{y}_2||$ between
two vectors $\mathbf{y}_1=\mathbf{y}(t_1)$ and
$\mathbf{y}_2=\mathbf{y}(t_2)$ taken in the arbitrary moments of
time $t_1$ and $t_2$ is close to zero if and only if the distance
$S({\bf U}_1,{\bf U}_2)$ between two different states ${\bf
U}(x,t_1)$ and ${\bf U}(x,t_2)$ of the spatial extended system taken
in the same moments of time $t_1$ and $t_2$ is also small. The
distance $S({\bf U}_1,{\bf U}_2)$ has been defined as
\begin{equation}\label{s}
S({\bf U}_1,{\bf U}_2)=\left(\int\limits_0^1\left\|{\bf U_1}(
x,t)-{\bf U_2} (x,t)\right\|^2\, dx \right)^{1/2},
\end{equation}
where $||\cdot||$ is Euclidian norm.

According to the neighbour method it means that there is the
one-to-one correspondence between ${\bf U}(x,t)$ and
$\mathbf{y}(t)$, therefore we can use the constructed auxiliary
low dimensional system $\mathbf{y}(t)$ to find UPTSTs by means of
SD--method.

Having constructed the auxiliary flow system~(\ref{eq:FlowFromPDE})
we can use SD-method to detect UPOs in it and UPSTSs in the initial
spatial extended chaotic
system~(\ref{eq:PDE})--(\ref{eq:PDE_Puasson}), respectively. In
$\mathbb{R}^3$ space a plane $\rho(x=0.25,t)=1.0$ has been selected
as Poincar\'e secant. Let us denote the vectors
$\mathbf{y}(t_n)=(1,\rho(0.5,t_n),\rho(0.75,t_n))^T$ corresponding
to the $n$-th crossing the selected secant surface by the trajectory
$\mathbf{y}(t)$ as ${\bf y}_n$. Then the description of the system
dynamics can be made with the help of the discrete map
\begin{equation}\label{eq:Map}
 {\bf y}_{n+1}={\bf G}({\bf y}_{n}),
\end{equation}
where ${\bf G}(\cdot)$ is the evolution operator. Obviously, it is
impossible to find the analytical form for the operator $\bf G$, but
numerical integration of the initial system of partial differential
equations {(\ref{eq:PDE})--(\ref{eq:PDE_Puasson})} can give us a
sequence of values $\left\{{\bf y}\right\}_{n}$, generated by the
map~(\ref{eq:Map}).

\begin{figure}
\onefigure[scale=0.325]{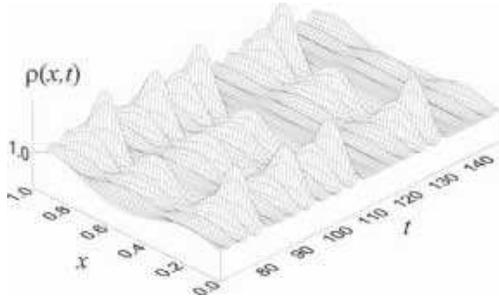} \caption{The spatio-temporal
dynamics of the charge density $\rho(x,t)$ of the electron beam of
Pierce diode. The oscillations for the selected control parameter
value $\alpha=2.858\pi$ are chaotic both in space and time.}
\label{fgr:Distribution}
\end{figure}

SD--method for picking out unstable periodic orbits in the
map~(\ref{eq:Map}) supposes consideration of the following map
\cite{Pingel:2001_SD-methodPRE}:
\begin{equation}\label{eq:SDMap}
 {\bf y}_{n+1}={\bf y}_{n}+\lambda {\bf C}\left[{\bf G}({\bf y}_{n})-{\bf
 y}_{n}\right],
\end{equation}
where $\lambda=0.1$ is the method constant and $\bf C$ is a
certain matrix of the set $\mathbf{C}_k$. Each of matrices $C_k$
should have only one non-vanishing entry $+1$ or $-1$ in row and
column, i.e., they are orthogonal. In two dimensions the complete
set of matrices consists of eight ones.

In works \cite{Schmelcher:1997_UnstableOrbit,
Pingel:2001_SD-methodPRE} it was shown that map~(\ref{eq:SDMap})
under the appropriate choice of the matrix $\bf C$ allows to
stabilize effectively the unstable saddle periodical orbits of
systems~(\ref{eq:Map}) and (\ref{eq:FlowFromPDE}). The positions
of the UPOs in  phase space are the same for the original chaotic
system~(\ref{eq:Map}) and the transformed dynamical
system~(\ref{eq:SDMap}) but their stability properties have
changed: unstable fixed points turned into stable ones. A
trajectory of transformed system~(\ref{eq:SDMap}) starting in the
domain of attraction of a stabilized fixed point  converges to it.
Therefore, the UPOs of a chaotic dynamical system~(\ref{eq:Map})
can be obtained by iterating the transformed
systems~(\ref{eq:SDMap}) using a robust set of initial conditions.
In our calculation the matrix
\begin{equation}\label{DP_Matrix}
    {\bf C} = \left(%
\begin{array}{cc}
  1 & 0 \\
  0 & 1 \\
\end{array}%
\right)
\end{equation}
is suitable to find UPOs in~(\ref{eq:Map}). Having obtained UPOs for
auxiliary systems~(\ref{eq:FlowFromPDE}) and (\ref{eq:Map}) we can
also obtain UPSTSs corresponding to them in the original spatial
extended system~(\ref{eq:PDE})--(\ref{eq:PDE_Puasson}).

The transformed system~(\ref{eq:SDMap}) allows to find only the
unstable periodic orbits of length $1$. To consider UPOs of length
$p$ the map
\begin{equation}
{\bf y}_{n+1}={\bf y}_{n}+\lambda {\bf C}\left[{\bf G}^p({\bf
y}_{n})-{\bf y}_{n}\right], \label{eq:SDMap-p}
\end{equation}
should be considered instead of~(\ref{eq:SDMap}) where ${\bf
G}^{(p)}(\cdot)$ is $p$-times iterated map~(\ref{eq:Map}). As far as
the spatial extended system and the auxiliary flow system are
considered, only the $p$-th crossing of the Poincar\'e secant by the
trajectory $\mathbf{y}(t)$ should be taken into account.

\begin{figure}
\onefigure[scale=0.35]{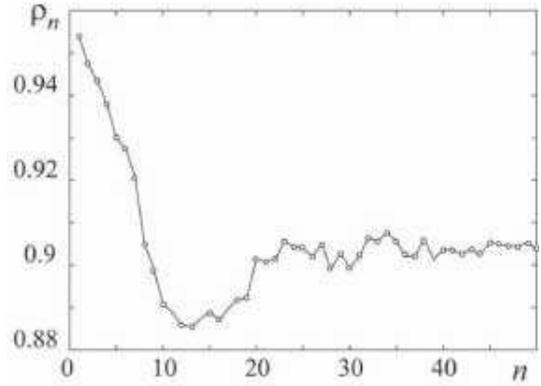} \caption{The dependence of the
space charge density $\rho_n(x=0.75)$ taken in the moments of time
when the trajectory $\mathbf{y}(t)$ in $\mathbb{R}^3$ space crosses
the Poincar\'e secant upon the number of iteration of SD--method for
the UPSTS of the length 1 ($T=4.2$). Pierce parameter has been
selected as $\alpha=2.858\pi$.} \label{fgr:Convergence}
\end{figure}

\begin{figure}
\onefigure[scale=0.5]{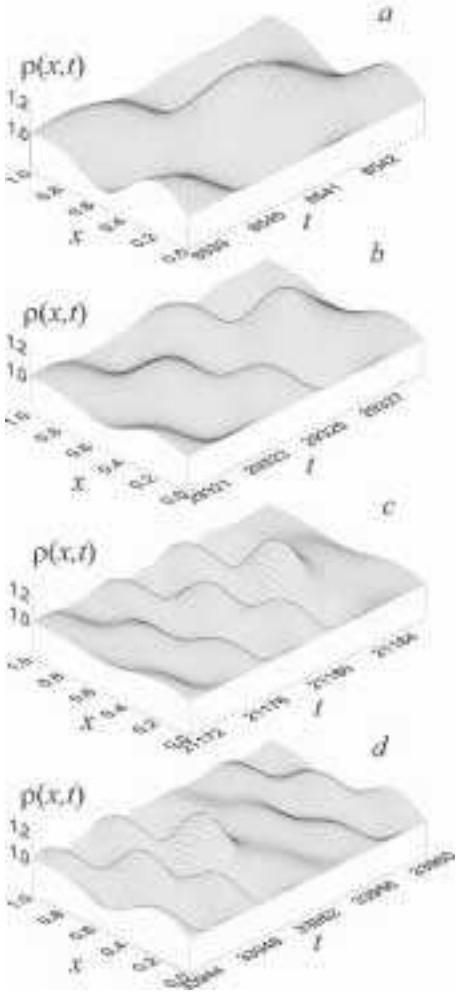} \caption{The distribution of space
charge density $\rho(x,t)$ corresponding to the unstable
spatio-temporal states with the following lengthes $p$ periods $T$:
(\textit{a}) $p=1$, $T=4.2$; (\textit{b}) $p=2$, $T=8.3$;
(\textit{c}) $p=3$, $T=16.9$; (\textit{d}) $p=4$, $T=18.9$. Pierce
parameter has been selected as $\alpha=2.858\pi$} \label{fgr:Orbits}
\end{figure}


So, by numerical iteration of the map (\ref{eq:SDMap-p}) with
different values of $p$ one can find the set of the unstable
periodic spatio-temporal states of the extended
system~(\ref{eq:PDE})--(\ref{eq:PDE_Puasson}). However, there is a
problem concerning with searching the state ${\bf U}(x,t_{n+1})$ at
the moment $t_{n+1}$ based on the known vector $\mathbf{y}_{n+1}$.
Indeed, we know only the coordinates of the state
$\mathbf{y}(t_{n+1})$ in the Poincar\'e secant but we don't know the
corresponding distribution of space charge density
$\rho(x,t_{n+1})$, velocity $v(x,t_{n+1})$ of the electron beam and
the potential $\varphi(x,t_{n+1})$, and, correspondingly, we do not
know the state $\mathbf{U}(x,t_{n+1})$ of the extended
system~(\ref{eq:PDE})--(\ref{eq:PDE_Puasson}). However, as we have
determined above with the help of the nearest neighbours method the
state $\mathbf{y}(t_{n+1})$ in the Poincar\'e secant uniquely
defines the corresponding spatial state $\mathbf{U}(x,t_{n+1})$
belonging to the attracting finite-dimensional subspace $W^s$ of the
infite-dimensional phase space $W^{\infty}$. To obtain this spatial
state $\mathbf{U}(x,t_{n+1})$ mentioned above we have used the
following procedure. The system of partial differential
equations~(\ref{eq:PDE})--(\ref{eq:PDE_Puasson}) describing the
fluid model of Pierce diode is integrated (and vector
$\mathbf{y}(t)$ is calculated) untill some vector ${\bf y}(t_s)$ is
close to the required one ${\bf y}_{n+1}$ with some demanded
precision: $||{\bf y}_{n+1}-{\bf y}(s)||<\delta$, where $\delta$ is
taken as $\delta=10^{-3}$. When this condition is satisfied, the
space state ${\bf U}(x,t_s)$ corresponding to the found vector ${\bf
y}(s)$ are considered as the required one ${\bf U}(x,t_{n+1})$ and
then the next iteration according to (\ref{eq:SDMap-p}) should be
done.

The spatio-temporal chaotic dynamics of the charge density
$\rho(x,t)$ of the electron beam of Pierce diode is shown in
Fig.~\ref{fgr:Distribution} for the Pierce parameter value
$\alpha=2.858\pi$. Applying the modified SD-method to the spatial
extended system allows to find the demanded periodical time-space
states.

The convergence of the iteration procedure~(\ref{eq:SDMap-p}) is
illustrated by Fig.~\ref{fgr:Convergence}, which shows the
dependence of the space charge density $\rho_n(x=0.75)$ in the
moments of time when the trajectory $\mathbf{y}(t)$ in
$\mathbb{R}^3$ space crosses the Poincar\'e secant upon the number
of iteration $n$ of the SD--method when the unstable periodic
spatio-temporal state of the length $p=1$ is studied. One can see
clearly that the iteration process of SD--method converges to the
value corresponding to the unstable time-periodical spatio-temporal
state of the system. Fig.~\ref{fgr:Orbits} shows the distribution of
space charge density $\rho(x,t)$ corresponding to the unstable
spatio-temporal states with different periods $T$ detected by means
of SD-method.

To verify both the correctness of the chosen value of the dimension
$m=3$ of the auxiliary system and the obtained results we have
repeated the SD-method procedure for the value of the auxiliary
system dimension being equal to $m=4$. In this case the
fourth-dimensional vector of the auxiliary is
\begin{equation}
{\mathbf{y}(t)=(\rho(0.2,t),\rho(0.4,t),\rho(0.6,t),\rho(0.8,t))^T}.
\label{eq:FlowFromPDE_m=4}
\end{equation}
For $m=4$ all UPSTSs founded coincide with the ones obtained above
for the dimension of the auxiliary system $m=3$, although the time
of calculations increases in this case sufficiently.

\begin{figure}
\onefigure[scale=0.35]{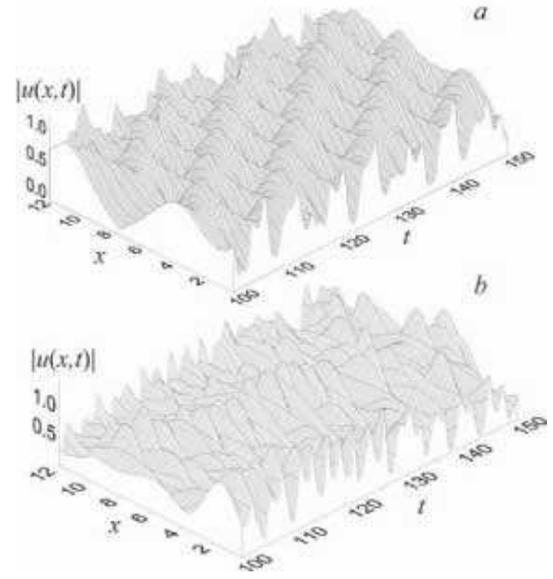} \caption{The spatio-temporal
dynamics of the Ginzburg-Landau equation. The evolution of
$|u(x,t)|$ is shown for the system length (\textit{a}) $L=12.63$ and
(\textit{b}) $L=13.25$} \label{fgr:CGLE_Distribution}
\end{figure}

To show the universality of the proposed approach we also report the
results of detecting the unstable periodic spatio-temporal states
for the one-dimensional complex Ginzburg-Landau equation (CGLE)
\cite{GIN-LAND:???_ModPhys}. The CGLE is a fundamental model for the
pattern formation and turbulence description. This equations is used
frequently to describe many different nonlinear phenomena in laser
physics \cite{Coullet:1989_Ginzburg-Landau}, chemical turbulence
\cite{Kuramoto:1981_Ginzburg-Landau}, fluid dynamics
\cite{Kolodner:1995_Ginzburg-Landau}, bluff body wakes
\cite{Leweke:1994_Ginzburg-Landau}, coupled spatial extended systems
\cite{Bragard:2000, Hramov:2005_GLEsPRE}.

\begin{figure}
\onefigure[scale=0.35]{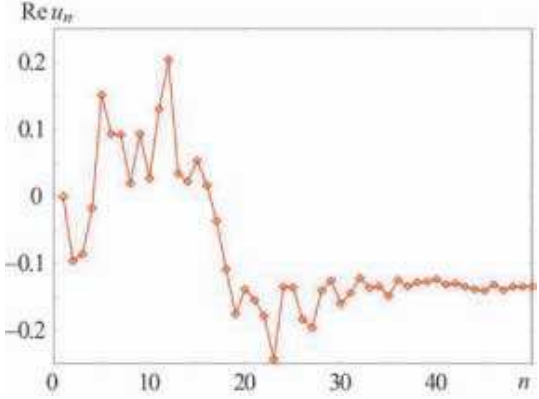} \caption{The dependence of the
value ${\rm Re}\,u_n(x=L)$ upon the number of iteration of
SD--method for UPSTS of the length $p=3$ ($T=20.2$). The values of
${\rm Re}\,u_n(x=L)$ are taken in the moments of time when the
trajectory $\mathbf{y}(t)$ in $\mathbb{R}^4$ space crosses the
Poincar\'e secant. System length has been selected as $L=13.25$.}
\label{fgr:CGLE_Convergence}
\end{figure}

We have considered one-dimensional CGLE
\begin{equation}
\frac{\partial u}{\partial t}=
u-(1-i\alpha)|u|^2u+(1+i\beta)\frac{\partial^2 u}{\partial x^2}
\label{eq:G-L}
\end{equation}
with periodical boundary conditions ${u(L,t)=u(0,t)}$. All
calculations were performed for a fixed system parameters
$\alpha=\beta=4$ and random initial conditions.  The numerical code
was based on a semi-implicit scheme in time with finite differences
in space. In all simulations we used a time step $\Delta t=0.0002$
for the integration and a space discretization $\Delta x=0.04$.

The system length $L$ has been chosen as the control parameter.
In our study we examined two values of the control parameter:
$L_1=12.63$ and $L_2=13.25$. For both these values of the control
parameter $L$ CGLE demonstrates the spatiotemporal chaotic regime.
The corresponding spatio-temporal chaotic dynamics of CGLE are shown
in Fig.~\ref{fgr:CGLE_Distribution} for the system lengths $L=12.63$
and $L=13.25$. One can see easily that the second case is
characterized by more complex irregular spatio-temporal chaotic
dynamics. Indeed, in the first case ($L=12.63$) the chaotic dynamics
is characterized by only one positive Lyapunov exponent
$\Lambda_1=0.04$, while the second chaotic regime ($L=12.63$) is
characterized by two positive Lyapunov exponents $\Lambda_1=0.10$
and $\Lambda_2=0.07$.

Applying the modified SD-method to the spatial extended CGLE we can
find the demanded unstable periodical spatio-temporal states as well
as for the fluid model of Pierce diode. We have constructed the
vector (\ref{eq:FlowFromPDE}) of the auxiliary low dimensional
system as
\begin{equation}
\mathbf{y}(t)=(u(x_1,t), \dots, u(x_m,t))^T,
\label{eq:FlowFromPDE_G-L}
\end{equation}
where $m$ is the dimension of the auxiliary system vector,
${x_i=iL/m}$, $i=\overline{1,m}$.

\begin{figure}
\onefigure[scale=0.4]{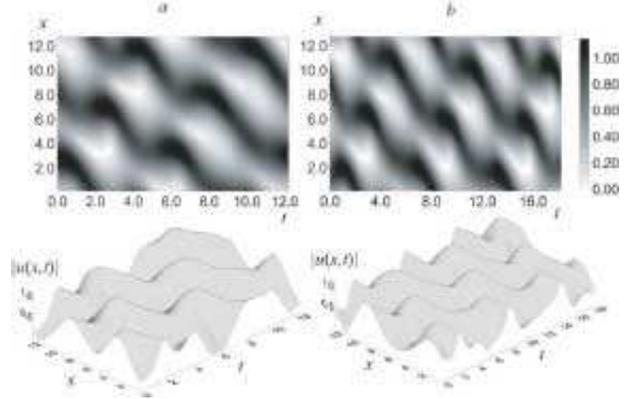} \caption{The evolution of the module
$|u(x,t)|$ corresponding to the unstable spatio-temporal states with
the following lengths $p$ and periods $T$: (\textit{a}) $p=1$,
$T=12.1$; (\textit{b}) $p=2$, $T=18.1$. System length has been
selected as $L=12.63$. The dimension of the auxiliary vector
$\mathbf{y}(t)$ has been chosen as $m=3$.}
\label{fgr:CGLE_Orbits_N=3}
\end{figure}

In contrast to the fluid model of Pierce
diode~(\ref{eq:PDE})--(\ref{eq:PDE_Puasson}) the dimension $m$ of
the auxiliary vector $\mathbf{y}(t)$ is unknown for
CGLE~(\ref{eq:G-L}). Therefore, we have to try to find UPSTSs by
means of the SD-method~(\ref{eq:SDMap-p}) for the different values
of the auxiliary system dimension $m$ starting from the minimal
dimension value $m=3$. If the required UPSTS is not found for the
selected value of the auxiliary vector dimension $m^*$, the
SD-method procedure should be repeated for the greater dimension
value $m=m^*+1$.

For the system length $L=12.63$ the dimension of the auxiliary
system $m=3$ is found to be adequate for the correct UPSTSs
detection. As it was mentioned above the system behavior is
characterized by one positive Lyapunov exponent. For the more
complicated case $L=13.25$ (when the behavior of CGLE is
characterized by two positive Lyapunov exponents) the dimension of
the auxiliary vector should be taken as $m=4$ for UPSTSs to be
detected successfully. The matrixes
\begin{equation}
\mathbf{C}=\left(
\begin{array}{rr}
-1 & 0\\
0 & -1\\
\end{array}
\right) \mbox{\quad and\quad} \mathbf{C}=\left(
\begin{array}{ccc}
1 & 0 & 0\\
0 & 1 & 0\\
0 & 0 & 1\\
\end{array}
\right)
\end{equation}
are found to be suitable to find UPSTSs for the system lengths
$L=12.63$ and $L=13.25$, respectively.


\begin{figure}
\onefigure[scale=0.375]{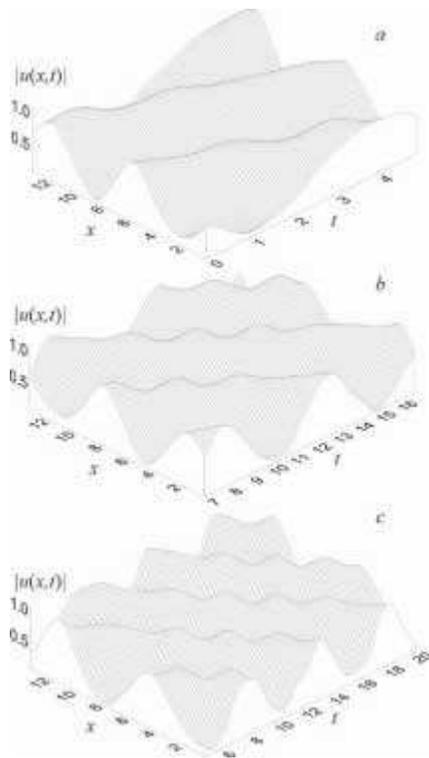} \caption{The evolution of the
proviles $|u(x,t)|$ corresponding to the unstable periodic
spatio-temporal states with the following lengths $p$ and periods
$T$: (\textit{a}) $p=1$, $T=4.95$; (\textit{b}) $p=2$, $T=16.83$;
(\textit{c}) $p=3$, $T=20.20$. The system length has been selected
as $L=13.25$. The dimension of the auxiliary vector $\mathbf{y}(t)$
has been chosen as $m=4$} \label{fgr:CGLE_Orbits_N=4}
\end{figure}

The convergence of the iteration procedure~(\ref{eq:SDMap-p}) is
illustrated in Fig.~\ref{fgr:CGLE_Convergence}. One can see clearly
that the iteration process of SD--method converges to the value
corresponding to the unstable periodic spatio-temporal state of the
system. Fig.~\ref{fgr:CGLE_Orbits_N=3} shows the evolution of the
profiles $|u(x,t)|$ corresponding to the unstable periodic
spatio-temporal states with the different periods $T$ detected by
means of SD-method for the system length $L=12.63$, when the
dimension of the auxiliary vector (\ref{eq:FlowFromPDE_G-L}) has
been chosen as $n=3$. The analogous evolution of the profiles
$|u(x,t)|$ corresponding to the unstable periodic spatio-temporal
states with the different lengths $p$ and periods $T$ is shown in
Fig.~\ref{fgr:CGLE_Orbits_N=4} for $L=13.25$ and $m=4$.

In conclusion, we have proposed the method of the detection of the
unstable periodic spatio-temporal states of spatial extended chaotic
systems being the extension of the well known SD-method. The
effectiveness of this method is illustrated by the consideration of
the fluid model of Pierce diode and the complex Ginzburg-Landau
equation.

\acknowledgments

We thank Dr Svetlana V. Eremina and Irene S. Rempen for the English
language support and the referees of our paper for the useful
comments. This work has been supported by U.S.~Civilian Research \&
Development Foundation for the Independent States of the Former
Soviet Union (CRDF, grant {REC--006}), Russian Foundation of Basic
Research (projects 06-02-72007-MNTI\_a, 05-02-16286 and
06-02-81013). We thank ``Dynasty'' Foundation. A.E.H. also
acknowledges support from the President Program, Grant No.
MD-1884.2007.2.


\begin{thebibliography}{0}

\bibitem{Cvitanovic:1988_cycles}
\Name{Cvitanovi\'c~P.} \REVIEW{Phys. Rev. Lett.}{61}{1988}{2729}.

\bibitem{Kostelich:1989_experiment}
\Name{Lathrop~D.~P. \and Kostelich~E.~J.} \REVIEW{Phys. Rev. A}
{40}{1989}{4028}.

\bibitem{Barreto:1997_bifurcations}
\Name{Barreto~E., Hunt~B.~R., Grebogi~C. \and  Yorke~J.~A.}
\REVIEW{Phys. Rev. Lett.}{78}{1997}{4561}.

\bibitem{Carroll:1999_UnstableOrbits}
\Name{Carroll~T.~L.} \REVIEW{Phys. Rev. E}{59}{1999}{1615}.

\bibitem{Cvitanovic:1991_orbits}
\Name{Cvitanovi\'c.} \REVIEW{Physica D}{51}{1991}{138}.

\bibitem{Rulkov:1996_SynchroCircuits}
\Name{Rulkov~N.~F} \REVIEW{Chaos}{6}{1996}{262}.

\bibitem{Pikovsky:1997_EyeletIntermitt}
\Name{Pikovsky~A.~S., Osipov~G.~V., Rosenblum~M.~G., Zaks~M. \and
Kurths~J.} \REVIEW{Phys. Rev. Lett.}{79}{1997}{47}.

\bibitem{Pikovsky:1997_PhaseSynchro_UPOs}
\Name{Pikovsky~A.~S., Zaks~M., Rosenblum~M.~G., Osipov~G.~V. \and
Kurths~J.} \REVIEW{Chaos}{7}{1997}{680}.

\bibitem{Aeh:2005_SpectralComponents}
\Name{Hramov~A.~E., Koronovskii~A.~A., Kurovskaya~M.~K., \and
Moskalenko~O.~I.} \REVIEW{Phys. Rev. E}{71}{2005}{056204}.

\bibitem{Rosenblum:1997_LagSynchro}
\Name{Rosenblum~M.~G., Pikovsky~A.~S. \and Kurths~J.}
\REVIEW{Phys. Rev. Lett.}{78}{1997}{4193}.

\bibitem{Pikovsky:1991_CSSymmetryBreaking}
\Name{Pikovsky A.~S. \and Grassberger P.} \REVIEW{J. Phys. A}
{24}{1991}{4587}.

\bibitem{Pazo:2002_UPOsSynchro}
\Name{Paz\'o~D., Zaks~M. \and Kurths~J.}
\REVIEW{Chaos}{13}{2002}{309}.

\bibitem{Bielawski:1993_ControllingChaos}
\Name{Bielawski~S., Derozier~D. \and Glorieux~P.} \REVIEW{Phys.
Rev. A}{47}{1993}{R2492}.

\bibitem{Ott:1990_ControllingChaos}
\Name{Ott~E., Grebogi~C. \and Yorke~J.~A.} \REVIEW{Phys. Rev.
Lett.}{64}{1990}{1196}.

\bibitem{Pyragas:1992_ControllingChaos}
\Name{Pyragas~K.} \REVIEW{Phys. Lett. A}{170}{1992}{421}.

\bibitem{Chen:1994_ControllingChaos}
\Name{Chen~Y.~H. \and Chou~M.~Y.} \REVIEW{Phys. Rev.
E.}{50}{1994}{2331}.

\bibitem{Pierre:1996_ControlChaosPlasma}
\Name{Pierre~Th., Bonhomme~G. \and Atipo~A.} \REVIEW{Phys. Rev.
Lett}{76}{1996}{2690}.

\bibitem{Franceschini:1999}
\Name{Franceschini~G., Bose~S. \and Sch\"oll~E.} \REVIEW{Phys.
Rev. E.}{60}{1999}{5426}.

\bibitem{Boccaletti:1999_ControllingChaos}
\Name{Boccaletti~S., Bragard~J. \and Arecchi~F.~T.} \REVIEW{Phys.
Rev. E.}{59}{1999}{6574}.

\bibitem{Rempen:2004_PierceDiode}
\Name{Rempen~I.~S. \and Hramov~A.~E.} \REVIEW{BRAS:
Physics}{68}{1998}{2004}{1998}.


\bibitem{Zoldi:1998_UPOs}
\Name{Zoldi S.~M. \and Greenside H.~S.} \REVIEW{Phys. Rev.
E}{57}{1998}{R2511}.

\bibitem{Schmelcher:1997_UnstableOrbit}
\Name{Schmelcher~P. \and Diakonos~F.~K.} \REVIEW{Phys. Rev.
Lett.}{79}{1997}{4734}.

\bibitem{Pingel:2001_SD-methodPRE}
\Name{Pingel~D., Schmelcher~P. \and Diakonos~F.~K.} \REVIEW{Phys.
Rev. E}{64}{2001}{026214}.

\bibitem{Godfrey:1987}
\Name{Godfrey~B.~B.} \REVIEW{Phys. Fluids}{30}{1987}{1553}.

\bibitem{Kuhn:1990}
\Name{Kuhn~S. \and Ender~A.} \REVIEW{J.Appl.Phys.}{68}{1990}.

\bibitem{Lindsay:1995}
\Name{Lindsay~P.~A., Chen~X. \and Xu~M.} \REVIEW{Int.
J.Electronics}{79}{1995}{237}.

\bibitem{Matsumoto:1996}
\Name{Matsumoto~H., Yokoyama~H. \and Summers~D.}
\REVIEW{Phys.Plasmas}{3}{1996}{177}.

\bibitem{Hramov:2004_IJE}
\Name{Hramov~A.~E. \and Rempen~I.~S.} \REVIEW{Int. J.Electronics},
{91}{2004}{1}.

\bibitem{Filatov:2006_PierceDiode_PLA}
\Name{Filatov~R.~A., Hramov~A.~E. \and Koronovskii~A.~A.}
\REVIEW{Phys. Lett. A}{358}{2006}{301}.

\bibitem{Pierce:1944}
\Name{Pierce~J.~R.} \REVIEW{\em J.Appl.Phys.}, 15:721, 1944.

\bibitem{Pecora:1995_statistics}
\Name{Pecora~L.~M., Carroll~T.~L. \and Heagy~J.~F.} \REVIEW{Phys.
Rev. E}{52}{1995}{3420}.



\bibitem{GIN-LAND:???_ModPhys}
\Name{Aranson I. S. \and Kramer L.} \REVIEW{Reviews of Modern
Physics}{74}{2002}{99}.

\bibitem{Coullet:1989_Ginzburg-Landau}
\Name{Coullet P., Gil P., \and Roca F.} \REVIEW{Opt.
Commun.}{73}{1989}{403}.

\bibitem{Kuramoto:1981_Ginzburg-Landau}
\Name{Kuramoto Y., \and Koga S.} \REVIEW{Prog.
  Theor. Phys. Suppl.}{66}{1981}{1081}.

\bibitem{Kolodner:1995_Ginzburg-Landau}
\Name{Kolodner P., Slimani S.,  Aubry N. \and  Lima R.}
\REVIEW{Physica D}{85}{1995}{165}.

\bibitem{Leweke:1994_Ginzburg-Landau}
\Name{Leweke T. \and Provansal M.} \REVIEW{Phys. Rev.
Lett.}{72}{1994}{3174}.


\bibitem{Bragard:2000}
\Name{Bragard J., Arecchi F. T. \and Boccaletti S.} \REVIEW{Int.
J. Bifurcation and Chaos}{10}{2000}{2381}.

\bibitem{Hramov:2005_GLEsPRE}
\Name{Hramov A. E., Koronovskii A. A. \and Popov, P. V.}
\REVIEW{Phys. Rev. E}{72}{2005}{037201}.


\end{thebibliography}
\end{document}